# Biodiversity analysis of metaproteomics samples with Unipept: a comprehensive tutorial

*Tim Van Den Bossche[1,2], Pieter Verschaffelt[1,2], Tibo Vande Moortele[3], Peter Dawyndt[3], Lennart Martens[1,2] & Bart Mesuere[3]*

(1) Department of Biomolecular Medicine, Ghent University, 9000, Ghent, Belgium
(2) VIB-UGent Center for Medical Biotechnology, VIB, 9000, Ghent, Belgium
(3) Department of Applied Mathematics, Computer Science and Statistics, Ghent University, Ghent, Belgium

**Corresponding author:** Lennart Martens

**Contact:** unipept@ugent.be

**ORCID:**

Tim Van Den Bossche: 0000-0002-5916-2587

Pieter Verschaffelt: 0000-0002-6675-1048

Tibo Vande Moortele: 0000-0002-7813-934X

Peter Dawyndt: 0000-0002-1623-9070

Lennart Martens: 0000-0003-4277-658X

Bart Mesuere: 0000-0003-0610-3441

**Abstract:**

Metaproteomics has become a crucial omics technology for studying microbiomes. In this area, the Unipept ecosystem, accessible at https://unipept.ugent.be, has emerged as an invaluable resource for analyzing metaproteomic data. It offers in-depth insights into both taxonomic distributions and functional characteristics of complex ecosystems. This tutorial explains essential concepts like Lowest Common Ancestor (LCA) determination and the handling of peptides with missed cleavages. It also provides a detailed, step-by-step guide on using the Unipept Web application and Unipept Desktop for thorough metaproteomics analyses. By integrating theoretical principles with practical methodologies, this tutorial empowers researchers with the essential knowledge and tools needed to fully utilize metaproteomics in their microbiome studies.

# 1. Introduction

Research into microbiomes and deciphering their functionality poses significant challenges. However, advancements in omics technologies are rendering this field increasingly accessible. These technologies facilitate an in-depth exploration of the functional biomolecular machinery. Among these advanced omics techniques, metaproteomics is emerging as a pivotal methodology, promising to greatly enhance our understanding of microbiome functionality by examining the spatio-temporal expression of microbial genes and their dynamics within a consortium of microorganisms *(1)*. This field has proven useful for investigating microbiomes across various environments, including the gut *(2)*, biogas plants *(3)*, soil *(4)*, and aquatic environments *(5)*. Despite these advances, a key challenge in the field remains: mapping identified peptides to proteins, and from there to taxa and functions. This challenge, often referred to as the protein inference problem, is particularly pronounced in metaproteomics because a given peptide may be found not only in multiple proteins or protein isoforms within the same species but also in homologous proteins from closely related species *(6, 7)*.

The Unipept ecosystem, available at https://unipept.ugent.be, addresses this challenge by bypassing the protein inference step and working directly with the identified peptides. Each tool in this ecosystem searches a list of tryptic peptides against the complete or part of the UniProtKB database *(8)*, after which a consensus taxon from the NCBI taxonomy *(9)* is assigned. Currently, five tools are deployed within the Unipept ecosystem: the web application *(10)*, the desktop application *(11, 12)*, the command line interface (CLI) *(13)* and the application programming interface (API) *(14)*. Additionally, there is the Unipept MetaGenomics Analysis Pipeline (UMGAP) *(15)* for metaproteomics analyses. Note that other (meta)proteomics tools such as MPA Portable *(16)* and PeptideShaker *(17)*, also have a direct export function to Unipept *(18)*.

# 2. Overview of the Unipept ecosystem for metaproteomics analysis

The Unipept ecosystem offers four distinct methods for metaproteomics analysis: the Unipept Web application, the CLI, the API and the desktop application (**Figure 1**).

The Unipept Web application is accessible via our website (https://unipept.ugent.be). It features an intuitive graphical user interface, making it the default gateway for most users. The web application offers interactive visualizations, intra- and inter-sample comparative analysis, and the ability to filter functional information by taxa. This application also supports advanced missed cleavage handling - or, in other words, which allows analyzing peptides with missed cleavages in contrast to the default option in Unipept which only analyzes (tryptic) peptides without missed cleavages (see **Section 4**). While user-friendly, the web application can only analyze relatively small metaproteomics samples with up to 25k peptides and has limited storage capacity.

The CLI is better suited for analyzing larger samples (over 25k peptides) and can be integrated into existing analysis pipelines. However, it lacks a graphical user interface and interactive visualizations, and does not currently support advanced missed cleavage handling. The API offers the flexibility to process large samples and allows other applications to incorporate Unipept's analysis features without requiring installation. Like the CLI, the API does not support advanced missed cleavage handling and requires significant programming knowledge to operate it.

The Unipept Desktop application is the newest addition to the Unipept ecosystem and combines features from the web application, CLI, and API into a single easy-to-use program. Despite requiring local installation, it offers all the advantages of the web application, such as an intuitive graphical user interface, interactive visualizations, intra- and inter-sample comparative analysis, the ability to filter functional information by taxa, and advanced missed cleavage handling. In addition, like the CLI and API, it can handle large files (up to 500k peptides), but can also handle custom protein databases - a feature unique to the Unipept Desktop application.

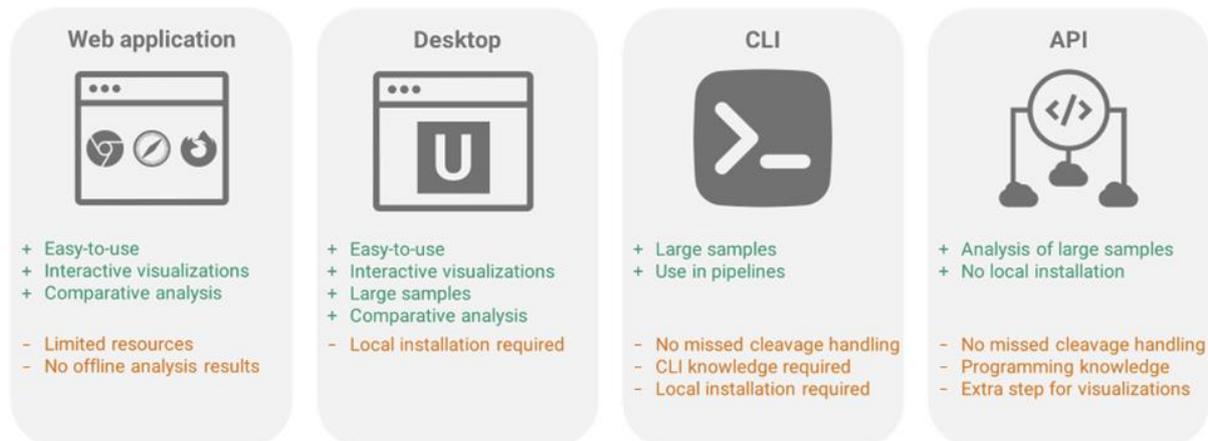

**Figure 1. Overview of the Unipept ecosystem for metaproteomics analysis with their advantages and disadvantages.**

In this comprehensive tutorial, we first discuss some general principles (LCA and advanced missed cleavage handling), after which both the web application and desktop stand-alone tool will be discussed. For the CLI, we refer to the 2018 CLI tutorial *(19)*, and for the API to our website (https://unipept.ugent.be/apidocs).

# 3. Unipept calculates the Lowest Common Ancestor for assigning taxa

By leveraging taxonomic information found in UniProtKB entries associated with a given tryptic peptide, Unipept uses a specialized algorithm to determine a consensus taxon for that peptide. This algorithm's key function is to calculate the Lowest Common Ancestor (LCA) among all taxa in which the peptide has been identified. The LCA is the most specific taxonomic level that encompasses all the taxa associated with a given peptide. For example, the peptide "TPAVFDMTK" can be found in eight different proteins assigned to seven different organisms, leading to the assignment of the family *Lachnospiracea* as the LCA (**Figure 2**).

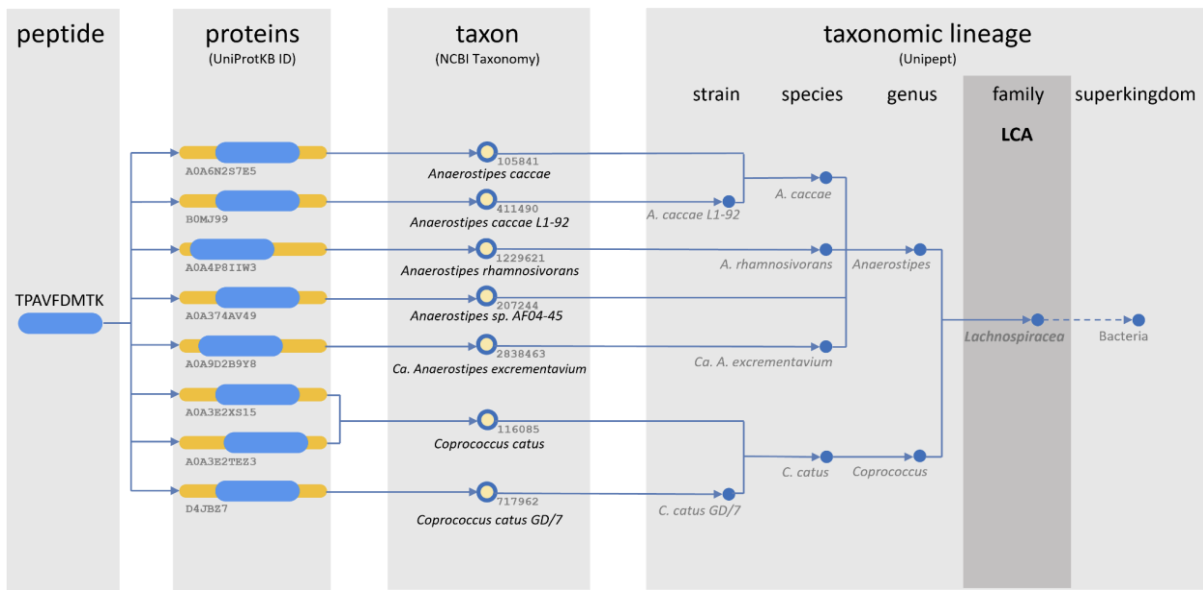

**Figure 2. Lowest Common Ancestor (LCA) calculation for a tryptic peptide.** In this figure, the peptide "TPAVFDMTK" can be found in eight different proteins assigned to seven different organisms. The LCA of these organisms is the family *Lachnospiracea*.

To perform the LCA calculation, Unipept relies on the Unipept taxonomy, a refined version of the NCBI Taxonomy. This is necessary, because, for example, NCBI Taxonomy ID 408170 is assigned to the species "human gut metagenome" (https://www.ncbi.nlm.nih.gov/Taxonomy/Browser/wwwtax.cgi?id=408170) - which can be considered a misclassification. Therefore, this cleaning process is essential, as the original taxonomy database might include taxonomic levels that are inconsistent or problematic. The Unipept Taxonomy uses a set of regular expressions to eliminate taxonomic errors, misclassifications, and inaccuracies with resilience, providing reliable results even when the original taxonomic information may be problematic.

# 4. Unipept's advanced missed cleavage handling

Enzymatic digestion of proteins with trypsin typically cleaves at the carboxyl (C-terminal) end of lysine (K) or arginine (R) residues, unless either residue is followed by a proline (P). However, missed cleavages may occur, where the enzyme fails to cleave at potential cleavage sites, resulting in longer peptides. Unipept's "advanced missed cleavage" accounts for these occurrences. In its default configuration, Unipept computes the LCA for only tryptic peptides. If a peptide contains missed cleavages, it will be *in silico* cleaved into multiple tryptic peptides. For instance, the peptide "TPAVFDMTKLAWMNGEYIK" with one missed cleavage, yields two tryptic peptides, each with its respective LCAs. The leading tryptic peptide "TPAVFDMTK" appears in eight UniProtKB entries (from seven unique organisms) with the *Lachnospiracea* family as its LCA (**Figure 2**). The trailing tryptic peptide, "LAWMNGEYIK," is present in twenty UniProtKB entries (from seventeen unique organisms), with superkingdom Bacteria as its LCA. However, longer peptides are more likely to be specific to a particular organism. Unipept's advanced handling of missed cleavages allows users to ascertain the LCA of complete peptides (including missed cleavages) by determining the LCA of the intersection of the organisms associated with each tryptic peptide. In the given example, this results in the LCA of four unique organisms, which is then attributed to the genus *Anaerostipes* (illustrated in **Figure 3**). Evidently, using longer peptides (with missed cleavages) leads to a more precise taxonomic annotation. However, employing this advanced missed cleavage function significantly impacts Unipept's performance. Therefore, It is recommended to use this function only in the Unipept Desktop application.

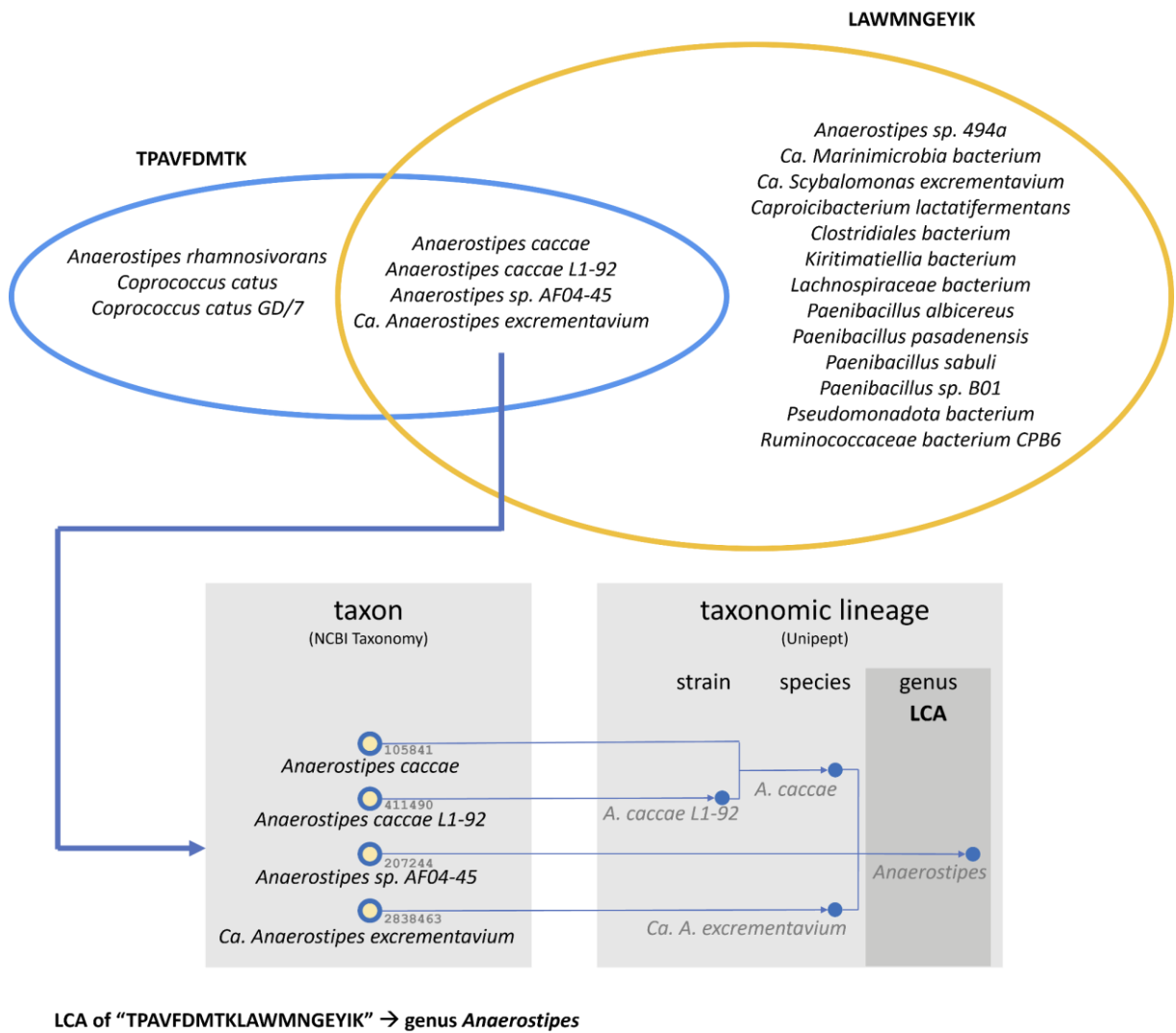

**Figure 3. Unipept's advanced missed cleavage handling.** For a peptide with missed cleavages, Unipept calculates the LCA for the organisms assigned to each tryptic peptide.

# 5. Unipept Web Application

The first tool we introduce here is the Unipept Web application, accessible at https://unipept.ugent.be. This user-friendly platform facilitates metaproteomic taxonomic and functional analysis for end-researchers, and presents the results through intuitive graphs and tables. Users can extract and export this information for use in other tools.

**Figure 4** showcases the Unipept homepage, including various tabs providing direct access to a variety of tools, including the "tryptic peptide analysis" tool (enabling the analysis of individual tryptic peptides) and the "metaproteomics analysis" tool for analyzing complete metaproteomic samples, containing thousands of tryptic peptides.

**Figure 4. The Unipept homepage.** The homepage is available on https://unipept.ugent.be. This page provides access to the different tools that are available, including thorough documentation.

## 5.1 Analysis of a single tryptic peptide

To understand Unipept's approach to summarizing taxonomic data linked to a peptide, we'll begin by analyzing an individual tryptic peptide. Click on "tryptic peptide analysis" in the banner to access the tool designed specifically for this purpose.

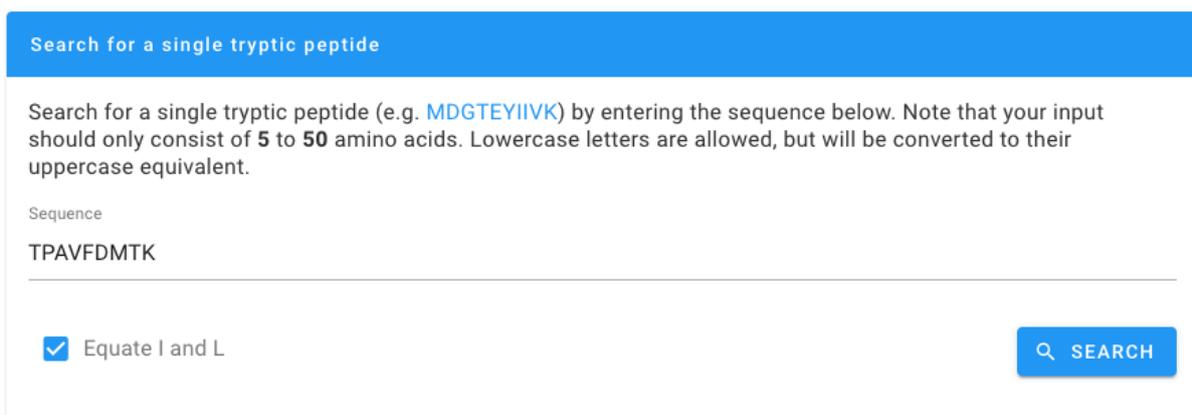

**Figure 5. The tryptic peptide analysis tool.**

Within this tool (as depicted in **Figure 5**), two inputs are required:

- **Sequence:** Enter the tryptic peptide sequence for analysis in this text field. For this tutorial, use **"TPAVFDMTK"** as an example. Copy and paste it into the text field.
- **Equate I and L:** Enabling this option causes Unipept to treat the amino acids isoleucine (I) and leucine (L) as equivalent. As the mass of these amino acids is identical, distinguishing between them is not possible. Users typically activate this option to prevent missing potential results. For instance, "MDGTEYIIVK**"** will also match with "MDGTEYLLVK**", "**MDGTEYILVK**"** and "MDGTEYLIVK**"** when this option is checked.

Click "search" or hit enter to begin the analysis. This process may take a few seconds.

**Figure 6. Overview of all proteins, taxa, and functional annotations associated with the provided tryptic peptide during the tryptic peptide analysis.**

Upon completion, a comprehensive table will present and summarize all the results. This page consists of two important parts: a tryptic peptide summary, and a graphical in-depth exploration of the results (**Figure 6**). The summary box details the findings related to the peptide, revealing matches with 6 distinct proteins in UniProtKB and establishing its LCA as the family *Lachnospiracea*. Alongside the taxonomic summary, the tool indicates associated functional annotations, such as Gene Ontology (GO) terms, Enzyme Commission (EC) numbers, and InterPro entries.

The second box, highlighted in green, offers an in-depth exploration of the analysis results. It provides several options to view different aspects of the peptide analysis results, enabling users to switch between six different views:

**i. Matched proteins**

This table lists every matched protein along with their associated taxonomic and functional annotations. Note that in this table, a single organism is referenced and not the aggregated LCA.

| UniProt ID | Name | Organism | Annotations |
|---|---|---|---|
| A0A6N2S7E5 | Glutamate--tRNA ligase | Anaerostipes caccae | GO EC IPR |
| B0MJ99 | Glutamate--tRNA ligase | Anaerostipes caccae L1-92 | GO EC IPR |
| A0A4P8IIW3 | Glutamate--tRNA ligase | Anaerostipes rhamnosivorans | GO EC IPR |
| A0A374AV49 | Glutamate--tRNA ligase | Anaerostipes sp. AF04-45 | GO EC IPR |
| A0A9D2B9Y8 | Glutamate--tRNA ligase | Candidatus Anaerostipes excrementavium | GO EC IPR |
| A0A3E2XS15 | Glutamate--tRNA ligase | Coprococcus catus | GO EC IPR |
| A0A3E2TEZ3 | Glutamate--tRNA ligase | Coprococcus catus | GO EC IPR |
| D4JBZ7 | Glutamate--tRNA ligase | Coprococcus catus GD/7 | GO EC IPR |

**Figure 7. Table listing every protein matched and its associated taxonomic and functional annotations.**

Based on the information from this table, you can see that eight different taxa match with the peptide (**Figure 7**). Unipept computes the LCA based on these eight taxon values, culminating in the family *Lachnospiracea* (**Figure 6**, box 1).

**ii. Lineage tree**

Under the lineage tree tab, a full-width tree view is presented, visually representing the different NCBI taxonomy ranks at each level (**Figure 8**). This visualization showcases the specific taxon associated with each rank concerning the particular tryptic peptide. When the taxa diverge at a specific level in the tree view, the nodes in the tree split into multiple child trees.

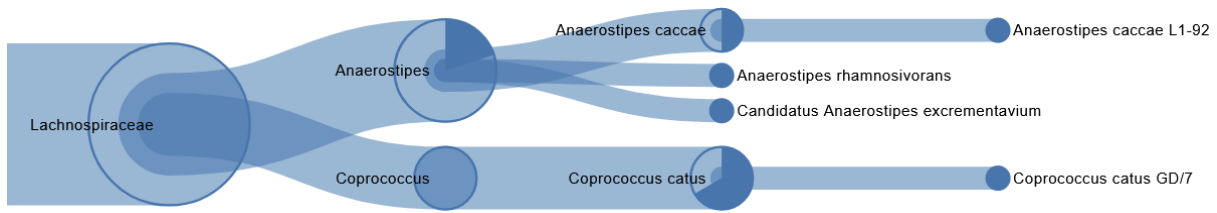

**Figure 8. Treeview visualization for a tryptic peptide.** In this particular example, all identified taxa correspond to *Lachnospiracea*. From this point forward, the identified taxa begin to diverge. *Lachnospiracea*, being the most specific taxon for all identified organisms, is reported as the Lowest Common Ancestor (LCA) for this peptide.

### iii. Lineage table

This table presents similar information to the lineage tree but in tabular format. It displays all known parent taxa at each corresponding rank for every identified taxon in the sample. In most cases, only certain parent taxa are known at various NCBI ranks, which explains why the example in **Figure 9** does not report values for 'kingdom', 'subkingdom', etc.

| Organism | superkingdom | kingdom | subkingdom | superphylum | phylum | subphylum |
|---|---|---|---|---|---|---|
| Anaerostipes caccae | Bacteria | | | | Bacillota | |
| Anaerostipes caccae L1-92 | Bacteria | | | | Bacillota | |
| Anaerostipes rhamnosivorans | Bacteria | | | | Bacillota | |
| Anaerostipes sp. AF04-45 | Bacteria | | | | Bacillota | |
| Candidatus Anaerostipes excrementavium | Bacteria | | | | Bacillota | |
| Coprococcus catus | Bacteria | | | | Bacillota | |
| Coprococcus catus | Bacteria | | | | Bacillota | |
| Coprococcus catus GD/7 | Bacteria | | | | Bacillota | |

Items per page: 10   1-8 of 8

**Figure 9. Lineage table when analyzing the tryptic peptide "TPAVFDMTK".**

### iv. Gene Ontology terms

Generates a list of all GO terms that are associated with the provided tryptic peptide and indicates the number of matched proteins containing these terms (**Figure 10**). The relationship of the different terms in each GO domain is then visually presented as a graph. Note that we also developed MegaGO *(20)*, a user-friendly tool that relies on semantic similarity between GO terms to compute the functional similarity between multiple data sets. However, discussing this tool in-depth falls out of scope of this tutorial.

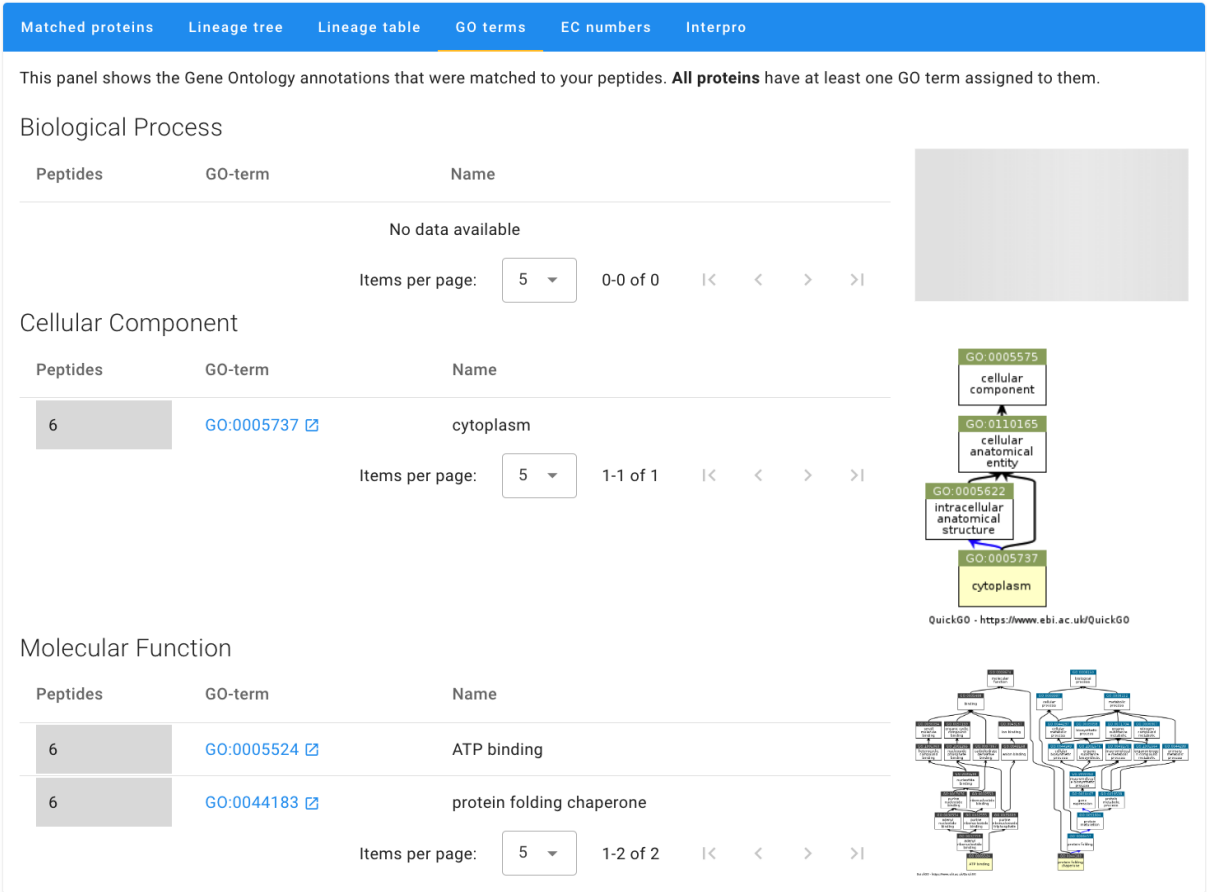

**Figure 10. Gene Ontology terms associated with the given tryptic peptide, presented in a tabular format.** A separate table is created for each of the three different Gene Ontology domains: biological process, cellular component, and molecular function.

**v. Enzyme Commission numbers**

Produces a list of all EC numbers associated with the provided tryptic peptide, similar to the presentation of GO terms (**Figure 11**).

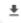

**Figure 11. Enzyme Commission numbers terms associated with the given tryptic peptide.**

### vi. InterPro

Renders a listing of all InterPro entries that are associated with the provided tryptic peptide, in a similar fashion as the GO terms and EC numbers (**Figure 12**).

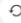

**Figure 12. InterPro entries associated with the given tryptic peptide.**

For this tutorial, it's recommended to familiarize yourself with this data exploration tool and comprehend the information provided, relating it to the underlying functionality of Unipept.

## 5.2 Metaproteomics analysis

### 5.2.1 Basic analysis of a metaproteomics sample

After exploring the analysis of individual tryptic peptides, we now progress to examining a complete metaproteomics sample, encompassing thousands of peptides. In this tutorial, we will use peptide lists obtained from the Critical Assessment of MetaProteome Investigation (CAMPI) study, the first community-driven, multi-laboratory comparison in metaproteomics. In this benchmark study, organized by the Metaproteomics Initiative *(21)*. In this study, two samples were used: a lab-assembled mixture simulating the human gut (SIHUMIx) *(22)*, and a fecal sample. In this section, we'll use a peptide list containing 25k randomly selected peptides from the "S11_PS_REF" dataset obtained from analyzing the SIHUMIx sample, here abbreviated "S11". For your convenience, this file can be downloaded from https://unipept.ugent.be/tutorial/S11.txt.

Start the metaproteomics analysis by clicking on "Metaproteomics Analysis" from the Unipept top menu. This will lead you to the page displayed in **Figure 13**. Box 1 contains three settings: "Equate I and L", "Filter duplicate peptides" and "Advanced missed cleavage handling". The "Equate I and L" option enables Unipept to treat the amino acids I and L as equivalent, as explained in **Section 5.1**. The "Filter duplicate peptides" option removes peptides appearing more than once in the dataset. The "Advanced missed cleavage handling" option instructs Unipept to analyze peptides with missed cleavages as a single peptide (as explained in **Section 4**).

To start a metaproteomics analysis, we first need to provide a list of peptides to Unipept. Unipept currently supports four different input sources, which directly correspond with the four tabs in Box 2 of **Figure 13**.

- **Create:** This input source enables the direct pasting of a list of tryptic peptides into Unipept. The input format should present each peptide on a separate line, containing solely letters (excluding modification parameters or quantification metrics). Note that Unipept currently does not consider quantification metrics supplied by search engines, except for spectral counting when "Filter duplicate peptides" (Box 1, **Figure 13**) is disabled.
- **Sample data:** Unipept incorporates a selection of sample datasets coming from commonly studied environmental ecosystems, allowing users to directly explore the tool's functionalities.
- **PRIDE:** Metaproteomics samples available on the PRIDE repository *(23)* can be seamlessly imported by utilizing their PRIDE ID.
- **Local data:** Custom datasets loaded using the "Create" or "PRIDE" options, along with a designated dataset name, are automatically stored in Unipept for future access. Users can retrieve these previously created datasets from this tab.

**Figure 13. The home page of the metaproteomics analysis tool in Unipept.** Box 1: Selection of the analysis settings "Equate I and L", "Filter duplicate peptides" and "Advanced missed cleavage handling". Box 2: upload of peptide lists.

For this tutorial, we will use the default options. Download the peptide list from https://unipept.ugent.be/tutorial/S11.txt, insert the peptides into the "Create" tab, assign the dataset the name "S11," and click on "Add to selected datasets." Now, the dataset should become visible in the "selected datasets" box located on the left. Then, proceed by clicking

"search" to initiate the analysis. This action will automatically direct you to the analysis results page.

Processing all peptides from the input dataset may take a few seconds. Once the analysis is complete, the processed list of peptides and a variety of visualizations will appear on the page (**Figure 14**).

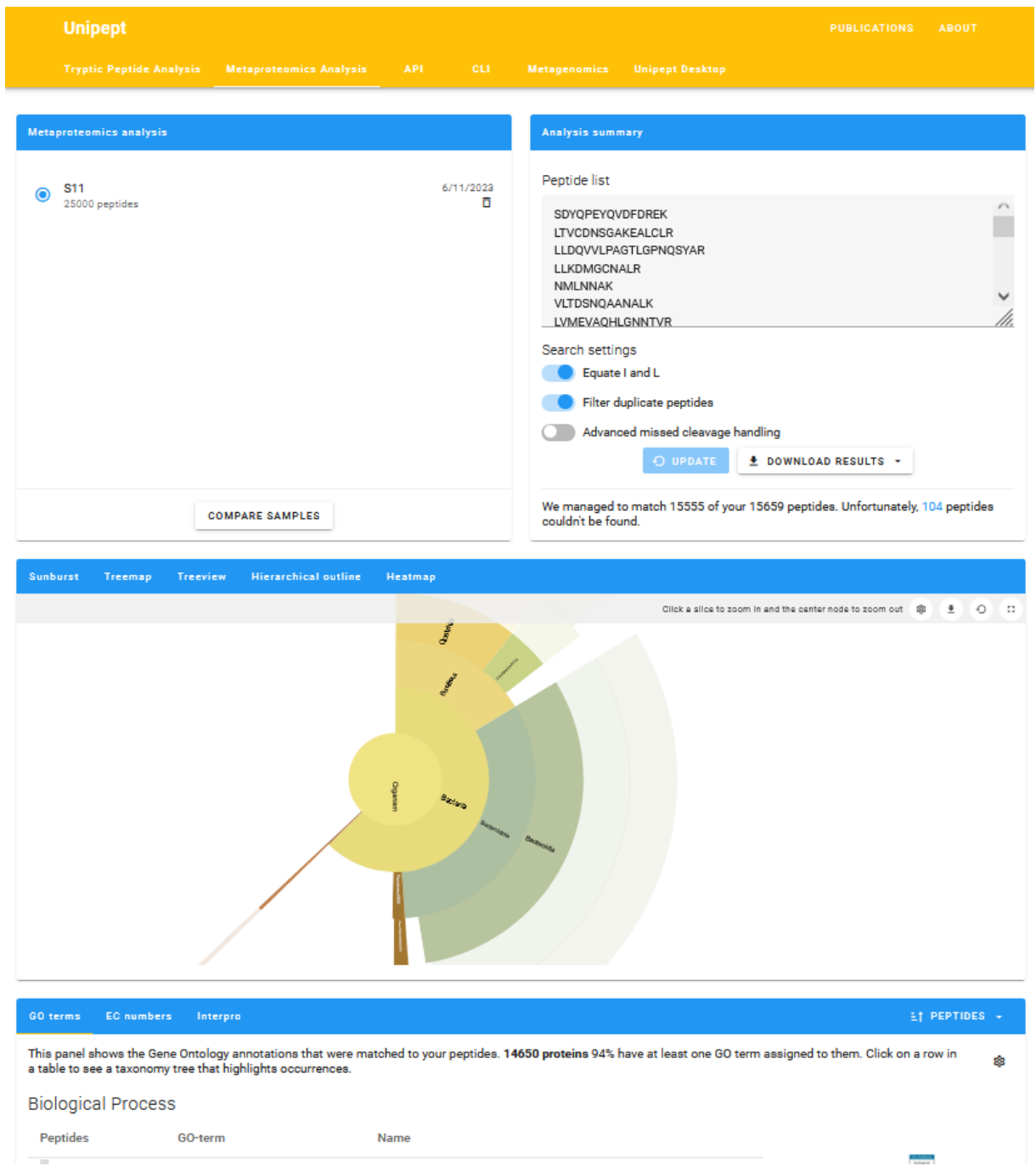

**Figure 14. Once the analysis is complete, the final results will appear on the page.**

The main results are split into a taxonomic and functional part. On top of the page, a summary of the analysis is displayed including some useful statistics, such as the amount of peptides that are found and the contents of the input dataset. Just below this summary is a card that shows all taxonomic analysis results and on the bottom is a third card that shows all functional analysis results.

**Taxonomic profile**

To investigate the taxonomic composition of a sample, Unipept provides four visualization options, each presenting the information in a unique manner. These visualizations offer individual users a deeper understanding of the organisms within the sample. The tabs on top of the taxonomic profile card allow users to switch between the different visualizations.

For this example, let's take a look at the Sunburst visualization (the primary display by default, **Figure 15**). This visualization enables a direct observation of the most abundant organisms in the sample.

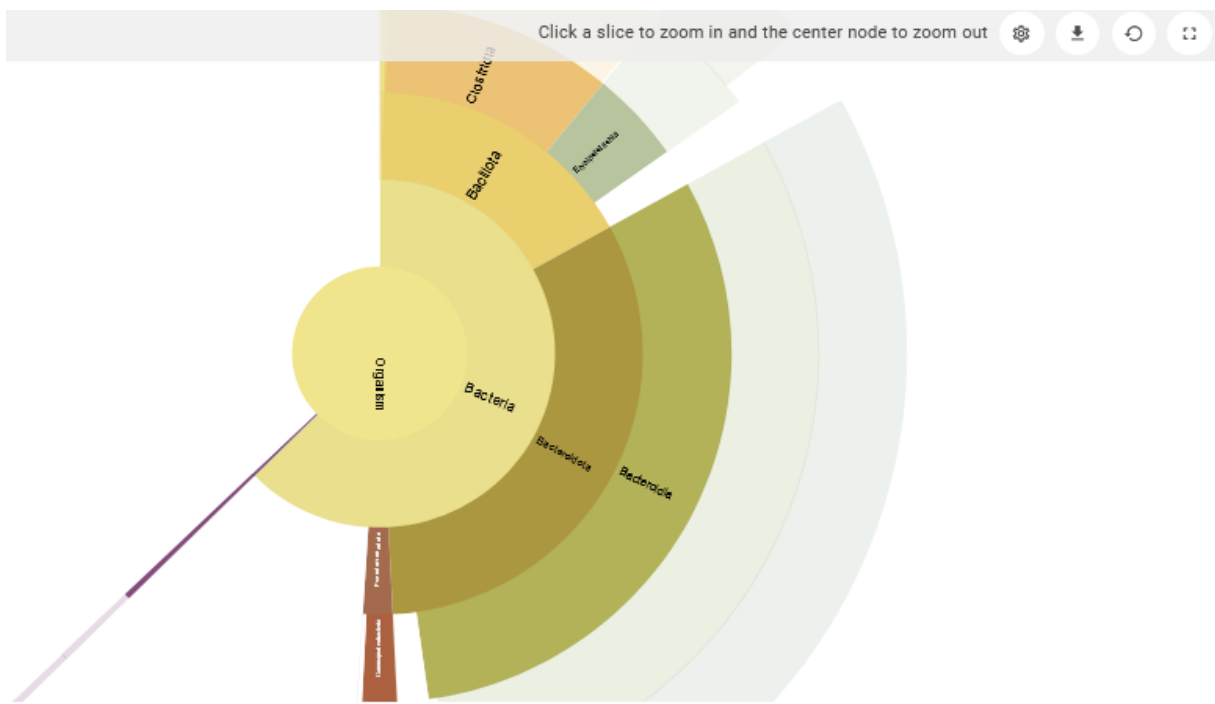

**Figure 15. The sunburst visualization of the example dataset "S11".**

Hovering the mouse cursor over different nodes in the sunburst visualization provides details about the number of peptides associated with an LCA corresponding to that specific node. For example, hovering over "Bacteria" reveals that 9740 peptides are specific to this level or lower, with 1909 peptides specific solely to this level, without further classification (i.e., with an LCA at the Bacteria superkingdom level).

From the 9740 - 1909 = 7831 peptides annotated with an LCA more specific than the Bacteria superkingdom level, we can observe that e.g. 5073 of them belong the phylum Bacteroidata, 2523 to the phylum Bacillota, and 225 to the phylum Pseudomonadota. All this information is accessible by merely hovering over different nodes in the Sunburst.

For deeper exploration of taxonomic information, clicking on a node corresponding to the LCA in the visualization automatically expands and rerenders the sunburst with the data pertaining to that NCBI taxon.

Make sure to explore the other available visualizations by switching between them using the tabs on top of the taxonomic profile card.

**Functional profile**

The functional profile of a sample can be explored in the last card on this page. A functional summary is provided for all identified GO terms, EC numbers and InterPro domains in a similar manner to the presentation in the tryptic peptide analysis tool. Users can switch between these different functional annotations using the tabs on top of the functional card.

**Exploring functions that are expressed by specific taxa**

Unipept offers a unique capability enabling the exploration of functions expressed by specific taxa identified in a sample. This functionality allows users to determine functions expressed by a subset of identified taxa. For example, if there's only an interest in functions expressed by bacteria, clicking the "Bacteria" node within one of the taxonomic visualizations will automatically update the tables with the functional annotations, displaying only those associated with bacteria.

Simultaneously, a new banner on top of the analysis results page will indicate the selected taxa, providing an option to reset the applied filter. This banner serves as a helpful navigational tool for users to manage the selected taxonomic filter during exploration.

### 5.2.2 Comparison of features within a single metaproteomics sample

The intra-dataset comparison component of the Unipept web application allows users to compare different features with each other, within a single dataset. This functionality is useful when users want to compare the relative abundance of organisms or functions within the dataset.

For this tutorial, we'll resume from the previous subsection, still using the S11 dataset. One of the tabs in the taxonomic visualization section of the Unipept results section is the "heatmap" tab. Click on this tab and a wizard will guide you through the heatmap creation process.

A heatmap always consists of two axes. For the single-dataset comparison, users have the liberty to choose which features they want to compare with each other, and thus require to select a set of features for each axis independently. The first step consists of selecting a data source for the horizontal axis. This can either be a collection of items that originate from the taxonomic or functional analysis. Select the "NCBI taxonomy" as data source and choose "species" as category. In the table below, a list of species appears in the table below that are present in the sample under investigation. All species are automatically sorted by number of peptides. In this particular example, select the first five species by ticking the respective boxes.

This will result in a selection similar to the one displayed in **Figure 16**. Proceed by clicking the "next" button to advance to the second step.

**Figure 16. Selection of the first five species from the sample under investigation for the horizontal axis of the heatmap.**

The next step involves the selection of the features for the second axis. It would be interesting to explore the functions predominantly expressed by the previously selected species. Therefore, choose "Enzyme Commission" as the data source for the vertical axis and select the five most abundant EC numbers. Click "next" to proceed.

The final configuration step in this wizard enables the selection of a normalization type to be applied to the data before generating the heatmap. Unipept supports three different normalization types:

- **All:** Computes the largest value over all cells in the heatmap and normalizes all cells based on this single value.
- **Rows:** Computes the largest value for each row in the heatmap and normalizes all cells on a row-by-row basis. Comparing relative values between different rows is not possible in this normalization, however, comparing relative values within one row is possible.
- **Columns:** Computes the largest value for each column in the heatmap and normalizes all cells on a column-by-column basis. Comparing relative values between different columns is not possible in this normalization, however, comparing relative values within one column is possible.

Select "All" for normalization type and click the "next". The heatmap will now appear on the screen, illustrating the comparison of different EC numbers across the five most abundant species found in the S11 dataset (**Figure 17**).

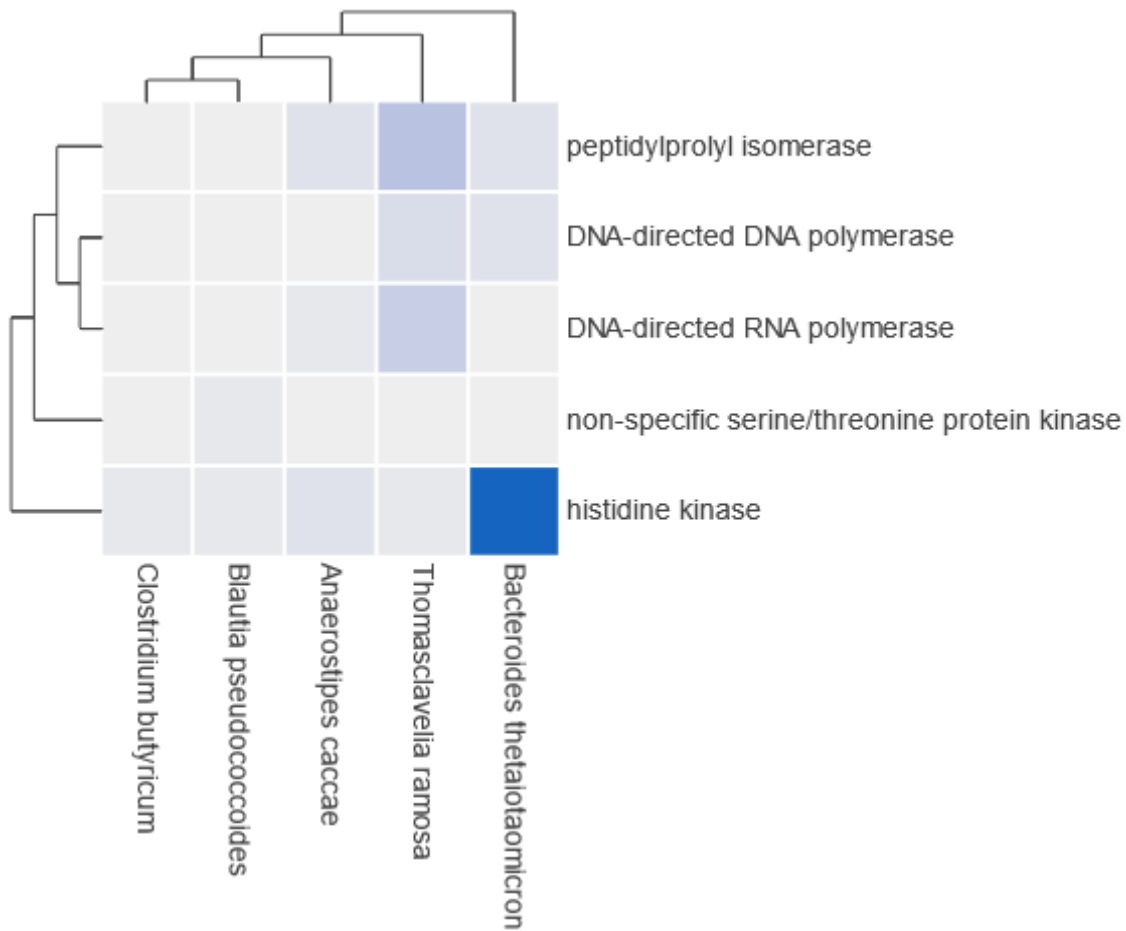

**Figure 17. Comparative heatmap of Enzyme Commission numbers across five most abundant species in the S11 dataset.**

### 5.2.3 Comparison of multiple metaproteomics samples

Up to this point, our focus has been on analyzing a single metaproteomics sample, but Unipept is also capable of comparing features between multiple datasets.

In contrast to selecting only one sample for analysis (as previously done), we will now choose four datasets for comparative analysis in Unipept. This approach allows for the construction of a new heatmap, which will compare the presence of ten species across these four datasets.

Start over and click on "Metaproteomics analysis" in the banner on top of the webpage to reset the application. For this experiment, we are going to randomly select 25k peptides from four datasets each from the CAMPI study *(24)*, all datasets originating from human fecal samples:

"F01_PS_MO", "F06_PS_MO", "F07_PS_MO", "F08_PS_MO" (abbreviated "F01", "F06", "F07", "F08", respectively). For your convenience, the files can be downloaded from: https://unipept.ugent.be/tutorial/F01.txt, https://unipept.ugent.be/tutorial/F06.txt, https://unipept.ugent.be/tutorial/F07.txt, and https://unipept.ugent.be/tutorial/F08.txt. Leave the configuration options to the default and click "search". As these samples involve more data to be processed, the analysis may take up a bit longer.

Once the analysis of all datasets is completed, click the white "Compare samples" button located at the bottom of the dataset selection card. Unlike the "intra-assay comparison" previously explored, in this comparison, you'll select the features for one axis of the heatmap, as the other axis is predefined by the datasets included in this analysis.

Choose "NCBI taxonomy" as a data source and set the category filter to "species". Select the first ten bacterial species and proceed by clicking "next". In the next step of the wizard, configure the normalization method for the final heatmap. For this analysis, choose to normalize over the columns. Click "next," and the resulting heatmap will be presented in the final step (**Figure 18**). This heatmap illustrates the relative abundance of ten selected species across the four different datasets. It highlights variations in species presence, indicating differences in their abundance within each dataset. The visualization allows for a quick comparison of how these species vary in their representation across the samples. For example, we observe that *Faecalibacterium prausnitzii* is relatively the most abundant species in samples F08 and F01, while *Ruminococcus bromii* is the relatively most abundant species in samples F07 and F06.

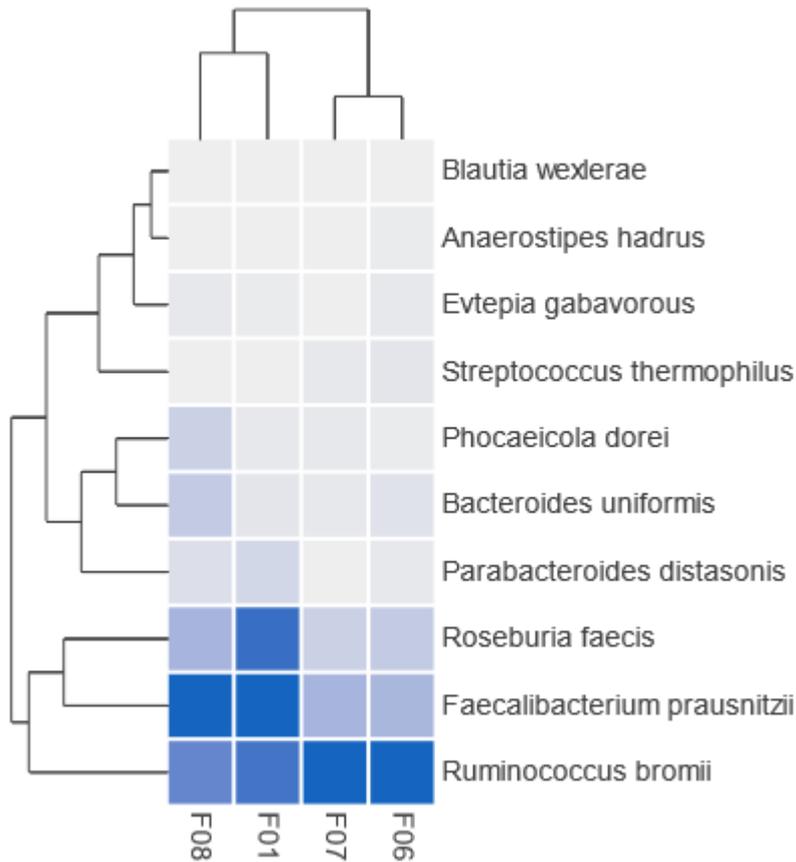

**Figure 18. Heatmap that compares the relative abundance of species over four datasets.**

# 6. Unipept Desktop

The Unipept Desktop, the most recent tool, combines features from the web app, CLI, and API into a user-friendly application. It facilitates handling large samples, making it suitable for researchers less acquainted with programming.

If you have used the Unipept Web application before, transitioning to the Unipept Desktop will be smooth, although there are a few distinct differences between the two. Operating as a standalone application, not a web-based platform, Unipept Desktop accommodates the analysis of significantly larger sample sizes, up to half a million peptides. Notably, it also introduces the capability to create custom protein reference databases, a critical feature and major improvement explored later in this tutorial.

## 6.1 Installation and getting started

Since Unipept Desktop is a standalone application, it requires installation and can be downloaded from [https://unipept.ugent.be/desktop#download](https://unipept.ugent.be/desktop#download). Download the installer that's most appropriate for your operating system and follow the on-screen instructions. If needed, you can also take a look at the in-depth installation guide available on our website that differentiates between the different major operating systems ([https://unipept.ugent.be/desktop](https://unipept.ugent.be/desktop)).

## 6.2 Metadata management and organization of datasets into studies

The concept of "projects" is unique to the Unipept Desktop app, enabling organization and storage of studies and assays. Projects are represented as folders on your hard drive, containing all related data, so analyses performed earlier can easily be reloaded and reused, as well as shared with other researchers.

### 6.2.1 Project structure

A project always consists of a collection of studies that, in turn, contain a collection of assays. Each of these terms is defined in a specific way:

- **Project:** A project is a collection of studies, and every study is a collection of assays. This terminology adheres to the definitions found in the ISA-tab standard ([https://isa-specs.readthedocs.io/en/latest/isatab.html](https://isa-specs.readthedocs.io/en/latest/isatab.html)).
- **Study:** A study is a collection of assays that are somehow related. They, for example, all originate from the same object under study. Each study must have a distinct name within a project; duplication of study names is not permitted and will be flagged as a conflict by the application.

- **Assay:** A study is a collection of assays that are somehow related. They, for example, all originate from the same object under study. Like studies, each assay must be uniquely named within the project to avoid naming conflicts flagged by the application.

### 6.2.2 Managing projects

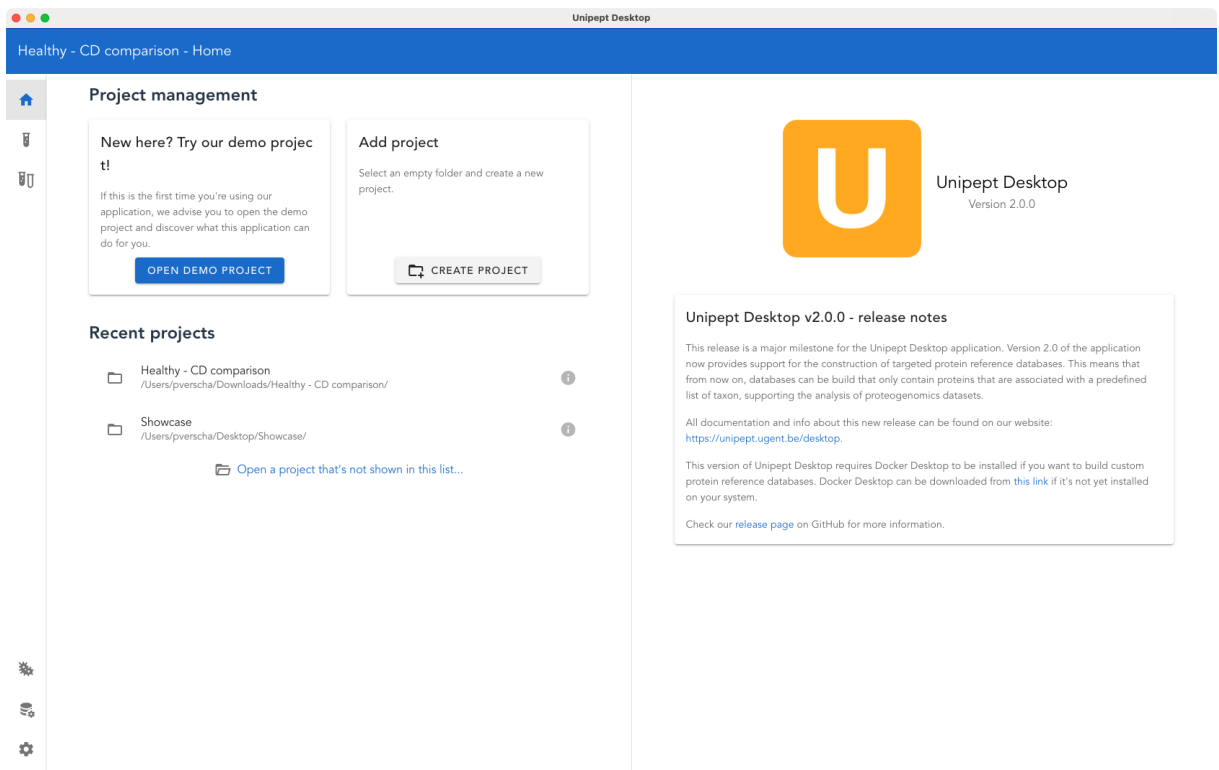

**Figure 19. The Unipept Desktop application start screen.** This screen shows the project management page as the primary interface.

The project management page (**Figure 19**) is the central hub for managing projects. It allows users to create new projects, open existing projects or delete old projects. The project management page is the first page that users see when they start the application and will also be shown whenever they click on the "home" button in the sidebar.

## 6.2.2 The demo project

For first-time users, it's recommended to begin by exploring the demo project to navigate through different features and pages. This demo project contains studies and pre-loaded assays, showcasing Unipept's capabilities. To access the demo project, click the "Open demo project" button on the home page.

## 6.2.3 Creating a project

To create a new project, click on the "Create new project" button on the home page. A dialog will open, allowing you to select an empty folder on your hard drive for the new project. If the selected folder already contains data, the application will prompt an error.

## 6.2.4 Opening a project

The home page always displays a list of the most recently created and opened projects. To open a project, click on one of the items in this list. If dealing with an old or externally sourced project, manually locate it on your hard drive by clicking "Open a project that's not shown in this list" and selecting the desired folder in the dialog.

## 6.2.5 Analyzing an assay

To add a new assay to the current project, ensure there's at least one study linked to the assay, because an assay must always be associated with a study. If no study exists, click on the "Add study" button in the project explorer to create one. A new "Study name" item will appear in the project explorer, representing a default name given to new studies, editable by double-clicking or via right-click and selecting "Rename study."

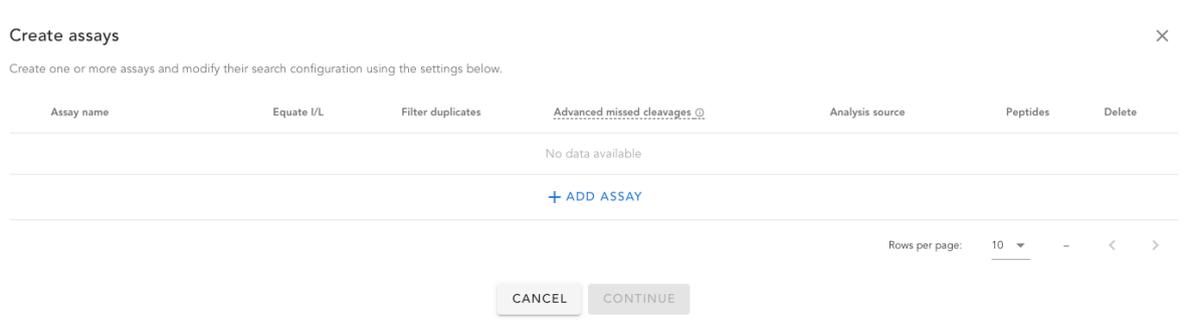

**Figure 20. The assay creation dialog.** This dialog guides you through the process of adding one or more assays to your project.

Once a study is created, adding a new assay is accomplished by clicking the file-and-plus icon beside the study name. A dialog will pop up (see **Figure 20**), allowing creation of one or more assays simultaneously. Clicking the "Add assay" button prompts a choice between two options.

The first option ("Empty assay") allows manual addition of a single assay. Paste a list of peptides into the text area and choose a suitable name for the new assay. The second option ("Bulk import from files") is suitable for importing files with many peptides from one or more files at once. Each input file generates a new assay entry with the assay names derived from the file names provided.

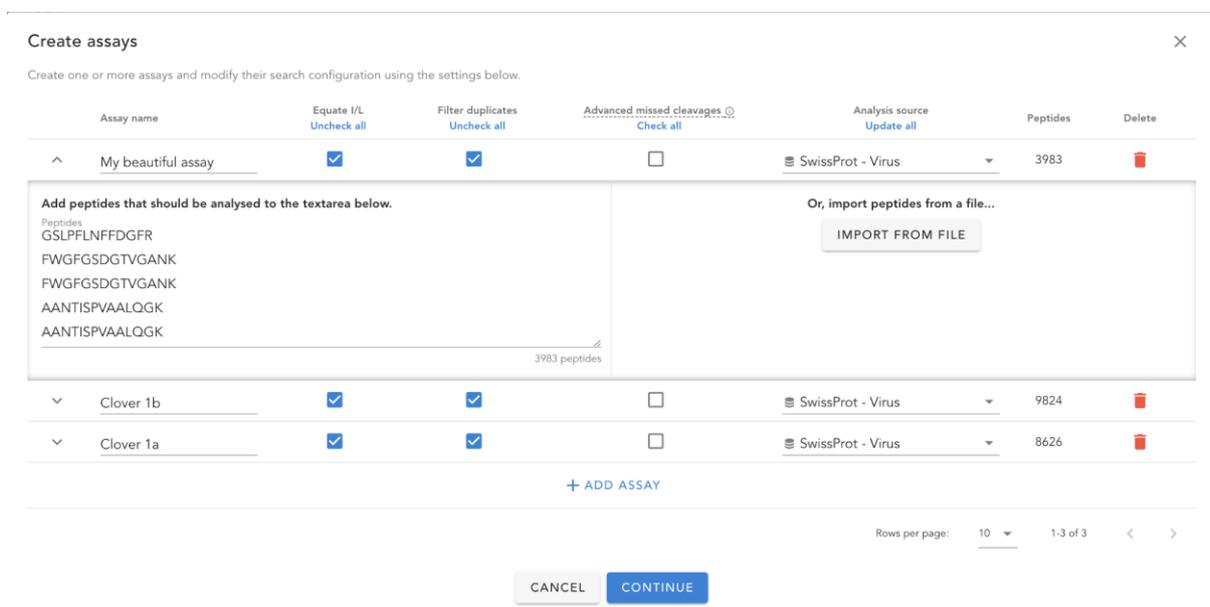

**Figure 21. Search settings for each of the new assays.** These settings can also be configured in the assay creation dialog.

It is important to note here that each assay can have its own search settings, impacting the metaproteomics analysis outcome. A new configuration option, the "Analysis source," is introduced, determining the protein reference database used for the sample analysis. The options include Unipept's online services (matching peptides against all proteins in UniProtKB, identical to the web application) or a local database (containing a subset of UniProtKB proteins). Local databases can be constructed on the "Targeted database management" page. The significance of these settings in the final analysis results will be explained in **Section 6.3**.

Finally, click on "continue" to import the assays into the current project and start the metaproteomic analysis. The analysis results page closely resembles that of the Unipept Web application, which you should already be familiar with at this point.

## 6.3 Constructing a custom protein reference database

The Unipept Desktop application's most significant improvement is enabling the construction of targeted or custom protein reference databases for metaproteomic analysis, which allows inferring the LCA at a more detailed level.

In the example from above (**Figure 2**), the peptide was chosen from the S11 dataset, analyzed in the CAMPI benchmark study *(24)* and in **Section 5.2**. The SIHUMIx sample *(22)* covers the most dominant phyla in human feces and consists of eight strains: the Firmicutes *Anaerostipes caccae DSMZ 14662, Clostridium butyricum DSMZ 10702, Erysipelatoclostridium ramosum DSMZ 1402* and *Lactobacillus plantarum DSMZ 20174*, the Actinobacteria *Bifidobacterium longum NCC 2705*, the Bacteroidetes *Bacteroides thetaiotaomicron DSM 2079*, the Lachnospiraceae *Blautia producta DSMZ 2950*, and the Proteobacteria *Escherichia coli MG1655*.

In this tutorial, we are going to construct a custom protein reference database for the SIHUMIx sample from scratch. Unipept Desktop contains a page especially dedicated to the creation and management of custom protein reference databases. Clicking the second from last bottom in the navigation bar on the left of the application (see **Figure 22**) will bring you to this management page.

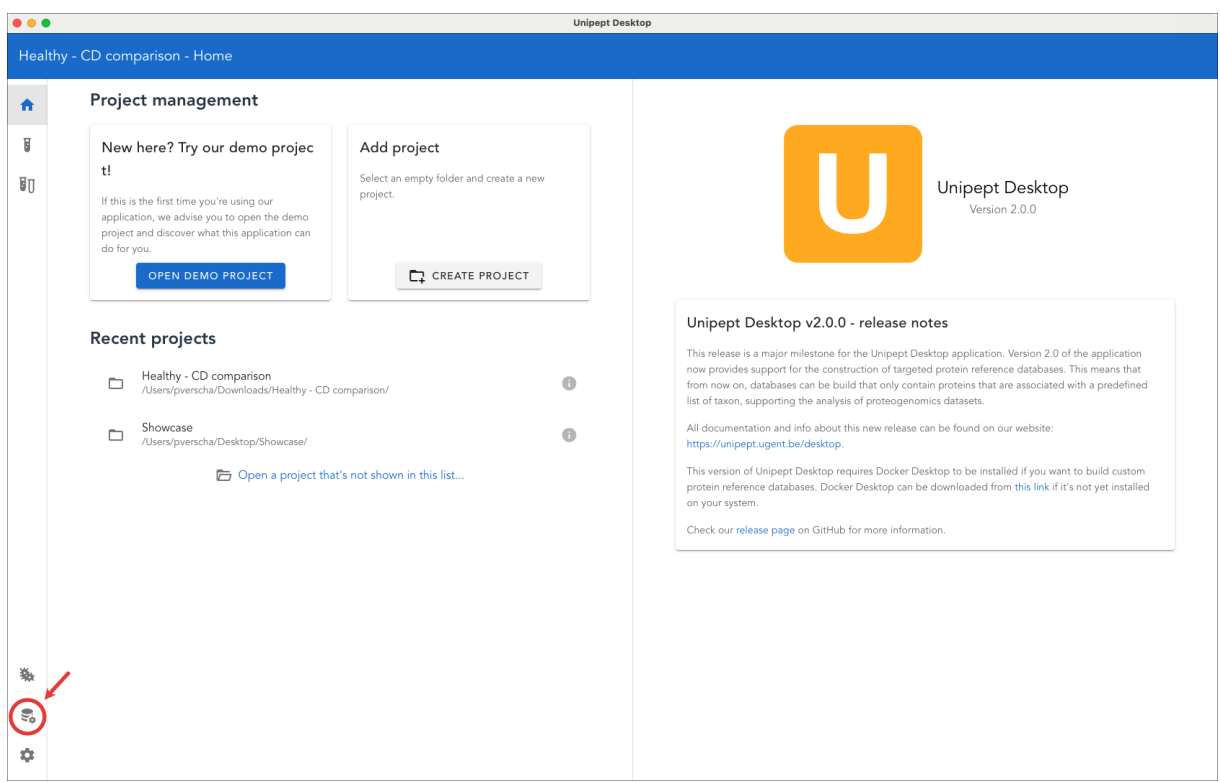

**Figure 22. By selecting the "custom database" icon, you can access the "custom database manager".**

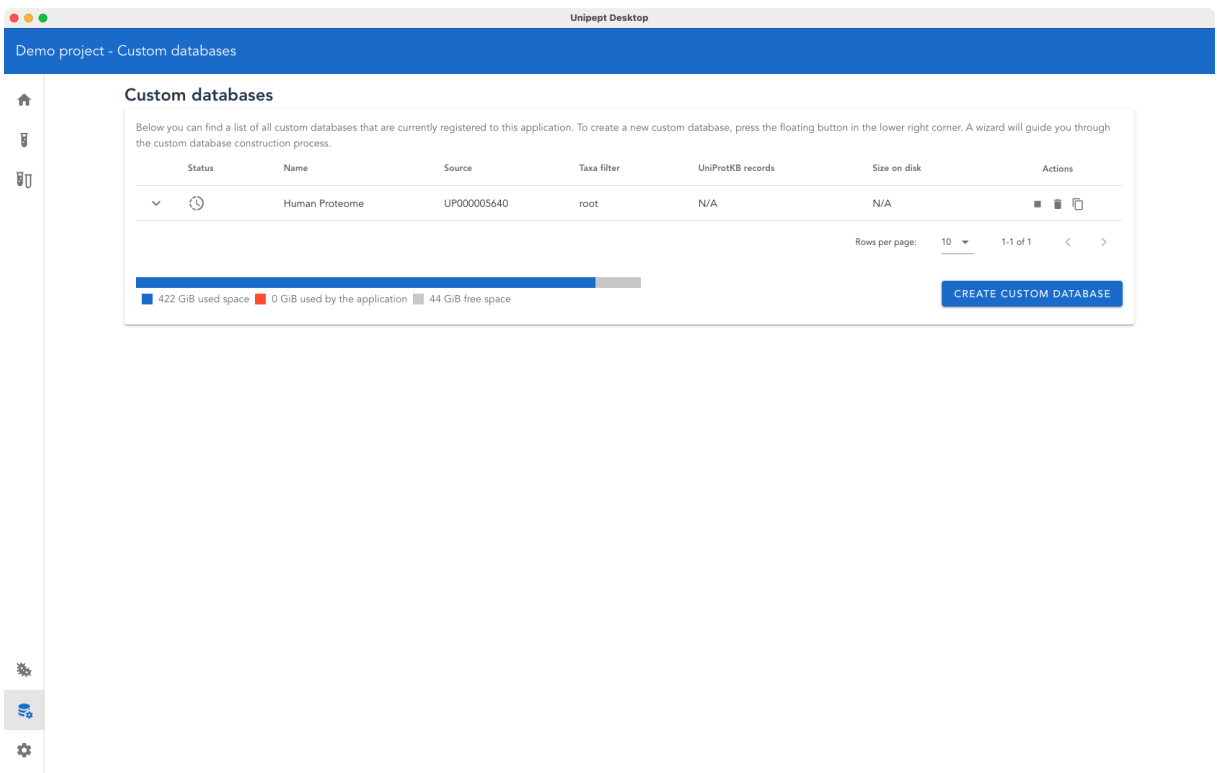

**Figure 23. Overview of the custom database manager**. The custom database manager shows a list of all databases present on your system.

Once the application brings you to the custom database manager, you'll see a list of databases present on your system (which will initially be empty). We have previously (in **Section 5.2**) performed the analysis of the SIHUMIx samples using the Unipept Web application and the default protein reference database that contains all sequences present in UniProtKB. For this tutorial, we are going to analyze the same samples using the Unipept Desktop application, but instead of trying to match them with a generalized protein reference database, we will be leveraging the fact that we have extra information about the ecosystem under study at our disposal. In real life experiments, extra potential taxonomic information can, for example, be obtained by first performing a metagenomics experiment.

To begin, click on the "Create Custom Database" button and proceed through the steps of the wizard. Provide a name for your custom database; we will be using "SIHUMIx". The wizard will present two options for database construction:

- **Manually filter database:** Manually select a list of taxa for which all associated proteins should be present in the custom reference database.
- **Custom reference protein:** Provide one or more UniProtKB reference proteomes which will integrally be used for database construction.

For our example, we are going to construct a database using all reference proteomes present for the organisms described earlier in the introduction, since this is the "prior knowledge" that we would like to exploit here. Click on the "custom reference protein" button and enter the following reference proteome identifiers: "UP000004935", "UP000515243", "UP000482084", "UP000000432", "UP000001414", "UP000515789", "UP000000625", and "UP000000439". By doing so, we are instructing Unipept to build a custom protein reference database that only contains the proteins that are directly associated with one of the organisms of these reference proteins.

**Figure 24. Selection of UniProtKB reference proteomes for custom database construction**.

After clicking "Build database", the construction process starts, and the progress is shown on screen. You can expand the current row to view more in-depth details about the ongoing operations (see **Figure 25**). Typically, the construction process completes within a few minutes.

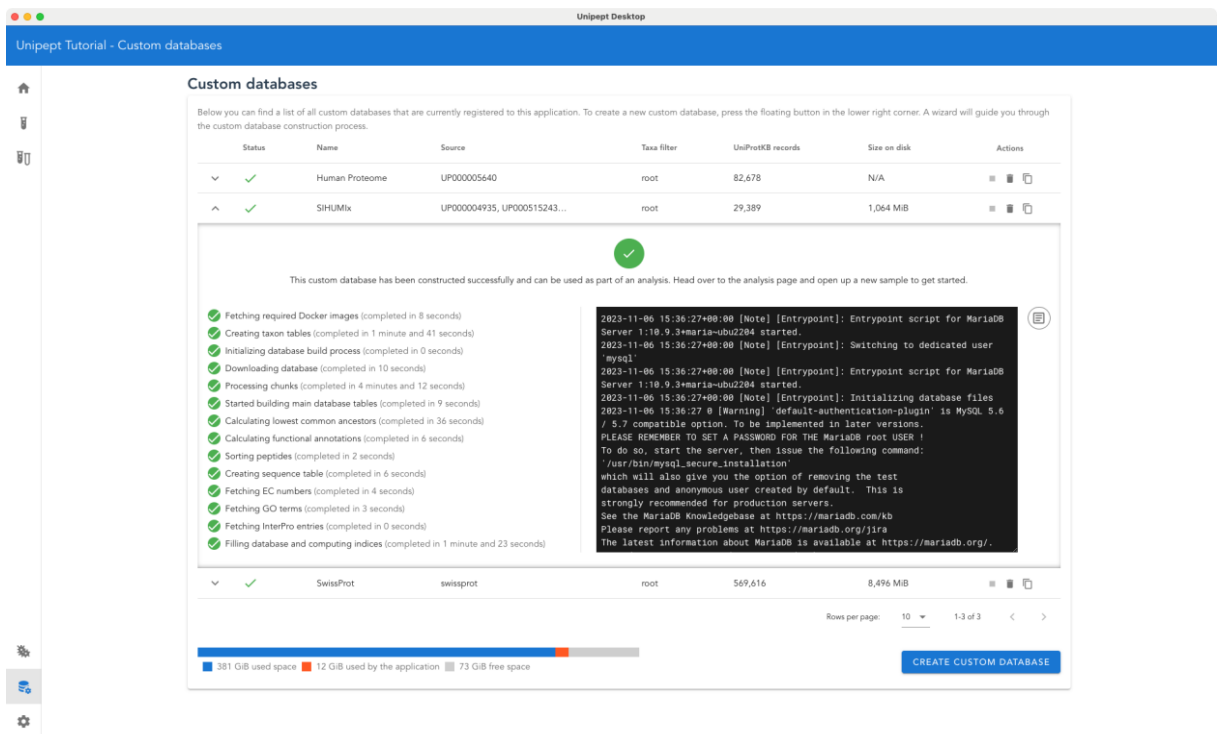

**Figure 25. Steps in the database construction process in Unipept Desktop.** Upon expanding the database row, an indicative timing for each step is shown, as well as the raw output that's being generated by Unipept under-the-hood. This output is not typically used by the end-user, but can be very useful to communicate to the developers in case of unexpected errors.

## 6.4 Re-analyzing SIHUMIx datasets using a custom protein reference database

In order to effectively see what the influence of the protein reference database used is on the analysis results, we are now going to use the database that we constructed in **Section 6.3** for the analysis of the SIHUMIx samples. Since there are less proteins present in our custom database, we can expect the amount of matches for each protein to also decrease, which should result in a more specific LCA annotation for some of the input peptides. For example, for peptide "TPAVFDMTK" which was analyzed in **Section 5.1** had six protein matches in the general protein reference database (UniProtKB) with an LCA assigned at the *Lachnospiracea* family. Now, by using a targeted protein database, only two proteins are found, leading to a more accurate LCA of the species *Anaerostipes caccae* (**Figure 26**).

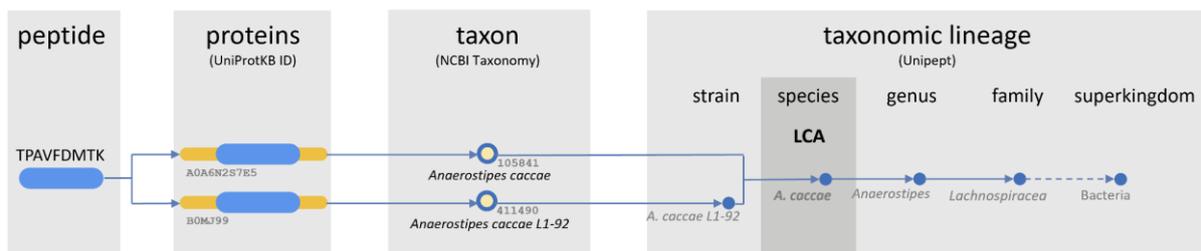

**Figure 26. Lowest Common Ancestor calculation with a customized database.** LCA shifts from family *Lachnospiracea* to species A*. caccae*.

### 6.4.1 Analyzing the SIHUMIx datasets using the general protein database

Start by creating a new project (as explained in **Section 6.2.3**). At this point, our objective is to find out what the impact is of the protein reference database on the final taxonomic analysis. Given that we're investigating two different protein databases, we will be creating two new studies, designated as "General reference DB" and "Custom DB". For this tutorial, we will be using six peptide lists from the SIHUMIx dataset from the CAMPI study. Each of these lists contains 25k peptides randomly chosen from their respective (complete) peptide lists. Please download these on:

- S03.txt (https://unipept.ugent.be/tutorial/S03.txt)
- S05.txt (https://unipept.ugent.be/tutorial/S05.txt)
- S07.txt (https://unipept.ugent.be/tutorial/S07.txt)
- S08.txt (https://unipept.ugent.be/tutorial/S08.txt)
- S11.txt (https://unipept.ugent.be/tutorial/S11.txt)
- S14.txt (https://unipept.ugent.be/tutorial/S14.txt)

We are going to import these samples into the application and analyze each of them twice - once using the general reference database and once using the custom SIHUMIx database. Starting with the general reference database, click the blue "Add new assay" button directly below the "General reference DB" study. The familiar wizard, as seen in **Section 6.2.5**, will appear. Choosing for "Bulk import from files" after clicking "Add assay" avoids the need to import each sample individually. Upon selecting this option, a file picker will open. Proceed to the directory where the files were downloaded, and select all files (**Figure 27**).

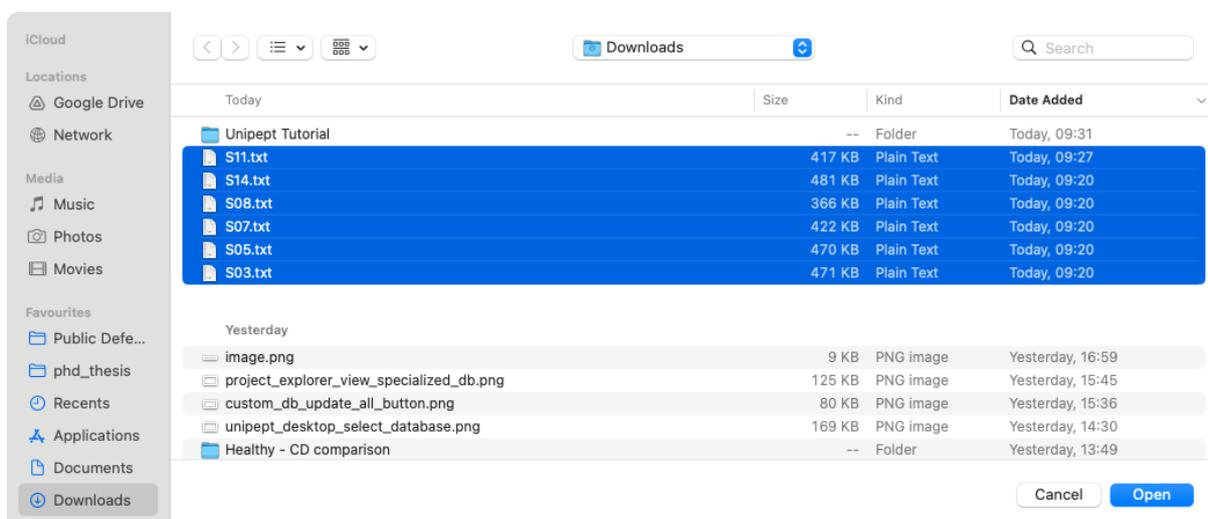

**Figure 27. Select all downloaded SIHUMIx peptide files.** You can use the shift or control keys to select multiple files at once.

After you've clicked "open", the files will be added to the list in the "assay creation wizard". Ensure that the analysis source is set to "Online service". This will prompt Unipept Desktop to use the generalized protein reference database hosted on our servers during the analysis of these assays. Click on "Continue" to start the analysis of these six samples using the default settings.

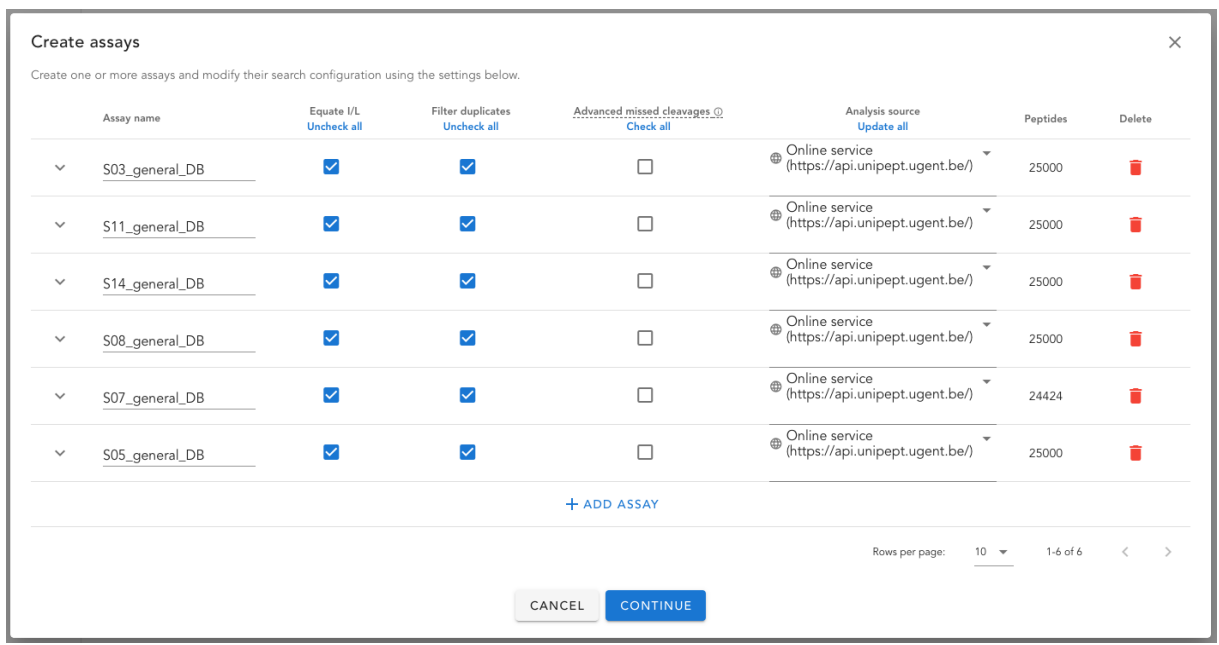

**Figure 28. After selecting the six datasets, use all default options.**

The analysis of these samples will take a few minutes. It's important to notice that we haven't yet used the custom protein reference database at this point. As a result, the taxonomic and functional analysis results of these assays should be identical to the ones produced by the Unipept Web application.

## 6.4.2 Analyze the SIHUMIx samples using the targeted protein database

We will resume from where we left off in **Section 6.4.1**, importing and analyzing the same set of SIHUMIx samples. However, instead of using Unipept's online service for the analysis, we will use the custom protein reference database we constructed in **Section 6.3**. Click the "Add new assay" button below the title of the second study ("Targeted DB"). Choose the same five files used previously, but ensure to change the default settings. To use our custom protein reference database, we must set it as the analysis source for each sample.

Click the "Update all" button below the "Analysis source" title (**Figure 29**). A small selection box will open, allowing you to select "SIHUMIx" among the options. Choose this database and click "Update all". Unipept Desktop will now use the custom protein database we created earlier for the analysis of these peptide samples instead of the online service.

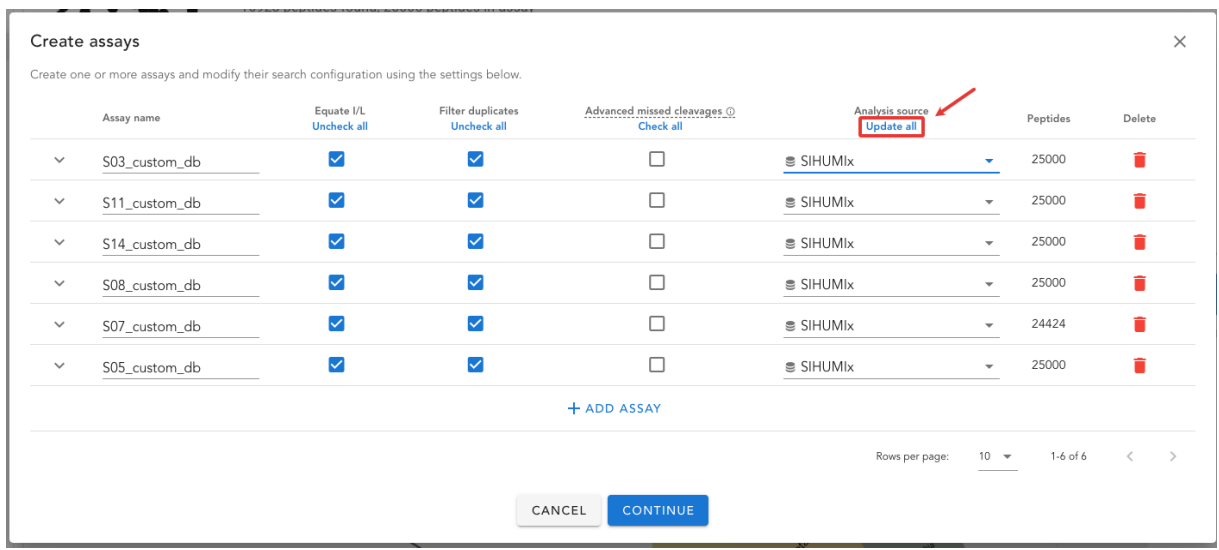

**Figure 29. Update the analysis source for all assays.** This enables you to change the analysis source to the targeted database that will be used for all assays.

After clicking continue, the samples will reappear to the project and two studies should now be visible in the project explorer located on the left side of the application, each study containing six samples (**Figure 30**).

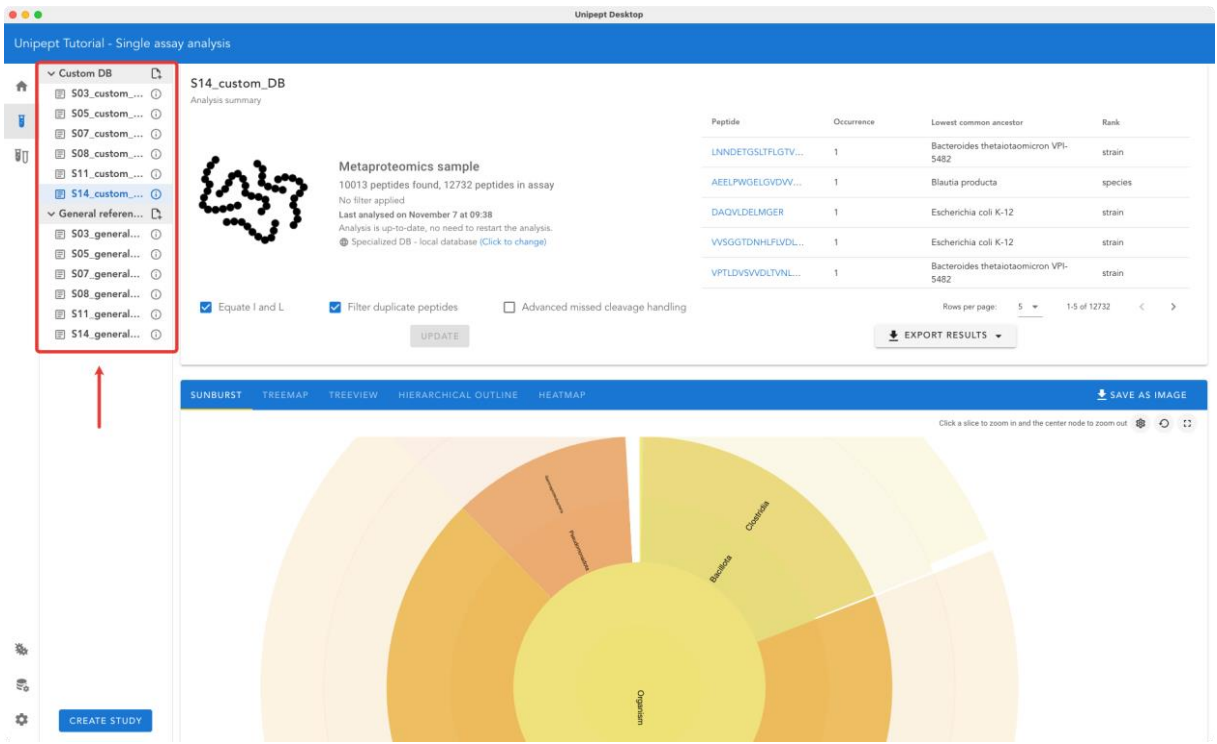

**Figure 30. All ten assays should be visible in the project explorer view on the left side of the application.**

6.4.3 Compare the impact of the protein reference database used on the analysis results

In the final part of this tutorial, we will explore the impact of using additional taxonomic information on the analysis results produced by Unipept Desktop. First, we're going to dive deeper into the taxonomic analysis performed on the S03 sample using both databases. When comparing the count of peptide matches between the targeted database and the general database for the S03 sample, there is a notable decrease in the count for the targeted database, as anticipated due to a higher possibility of random peptide matches or potential contaminants. However, the matches identified by the targeted database are much more specific.

Unipept Desktop allows users to export analysis results as CSV or TSV files, enabling their import into spreadsheet software. To do this, click on the "export results" button under the peptide summary table. The exported file contains data from this table, detailing each peptide's occurrences, LCA, and rank. **Figure 31** displays the total number of peptide matches found for sample S03 and categorizes them according to the major levels within the NCBI taxonomy. This figure is generated using taxonomic analysis information exported by Unipept Desktop.

**Figure 31** clearly indicates that the number of peptides annotated with an LCA at more specialized ranks of the NCBI taxonomy is significantly higher when using the targeted SIHUMIx database for the taxonomic analysis. This clearly showcases the impact of integrating supplementary, prior taxonomic information into Unipept's analysis.

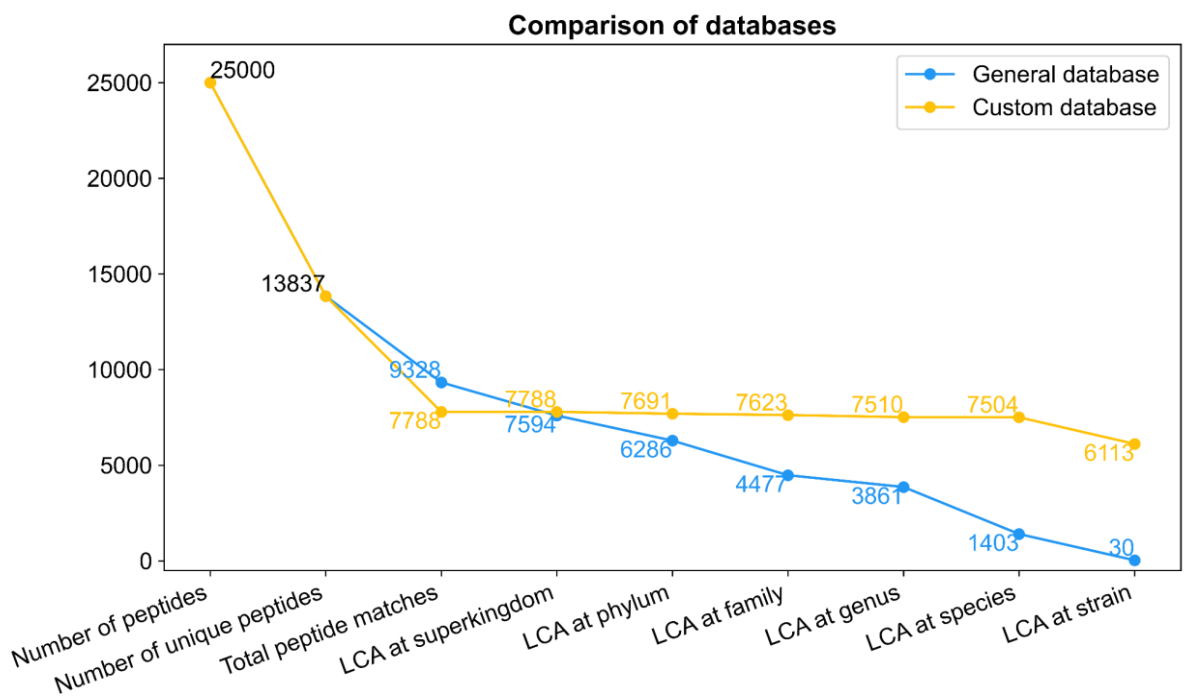

**Figure 31. A comparison of peptides matches for both databases with an LCA at the six major ranks in the NCBI taxonomy.** While the total count of peptide matches is higher when using the general UniProtKB protein database for analysis (blue line, general database), the count drastically changes when examining more specialized NCBI ranks, particularly species or strain, which are of significant interest to most researchers (yellow line, custom database).

# 6. Acknowledgements

This work has benefited from collaborations facilitated by the Metaproteomics Initiative (https://metaproteomics.org/) whose goals are to promote, improve and standardize metaproteomics *(21)*. Part of this work was supported by the Research Foundation — Flanders (FWO) for ELIXIR Belgium (I002819N). T.V.D.B. acknowledges funding from the Research Foundation Flanders (FWO) [1286824N]. L.M. acknowledges funding from the European Union's Horizon 2020 Programme (H2020-INFRAIA-2018-1) [823839], from the Research Foundation Flanders (FWO) [G028821N][G010023N][W001120N] and from Ghent University Concerted Research Action [BOF21/GOA/033] and from an ELIXIR Implementation study.

# 7. Ethics declarations

The authors declare no competing interests.